\makeatletter
\disable@package@load{program}{}
\makeatother

\documentclass[pdflatex,Namedate,sn-basic,iicol]{sn-jnl} 

\usepackage{aas_macros}

\usepackage{graphicx}%
\usepackage{multirow}%
\usepackage{amsmath,amssymb,amsfonts}%
\usepackage{amsthm}%
\usepackage{mathrsfs}%
\usepackage[title]{appendix}%
\usepackage{xcolor}%
\usepackage{textcomp}%
\usepackage{manyfoot}%
\usepackage{booktabs}%
\usepackage{algorithm}%
\usepackage{algorithmicx}%
\usepackage{algpseudocode}%
\usepackage{listings}%

\usepackage{orcidlink}%

\begin{document}

\title[Article Title]{
From stellar light to astrophysical insight: automating variable star research with machine learning
}


\author[1]{\fnm{Jeroen} \sur{Audenaert} \orcidlink{0000-0002-4371-3460}}\email{jeroena@mit.edu}

\affil[1]{\orgdiv{MIT Kavli Institute for Astrophysics and Space Research}, \orgname{Massachusetts Institute of Technology}, \orgaddress{\city{Cambridge}, \postcode{02139}, \state{MA}, \country{USA}}}


\abstract{
Large-scale photometric surveys are revolutionizing astronomy by delivering unprecedented amounts of data. The rich data sets from missions such as the NASA \textit{Kepler} and TESS satellites, and the upcoming ESA PLATO mission, are a treasure trove for stellar variability, asteroseismology and exoplanet studies. In order to unlock the full scientific potential of these massive data sets, automated data-driven methods are needed. In this review, I illustrate how machine learning is bringing asteroseismology toward an era of automated scientific discovery, covering the full cycle from data cleaning to variability classification and parameter inference, while highlighting the recent advances in representation learning, multimodal datasets and foundation models. This invited review offers a guide to the challenges and opportunities machine learning brings for stellar variability research and how it could help unlock new frontiers in time-domain astronomy.

}

\keywords{machine learning, asteroseismology, stellar variability}



\maketitle

\section{Introduction}\label{sec1}

The field of astronomy has undergone a major transformation in recent decades, driven by the arrival of large ground- and space-based surveys. The delivery of millions of high-precision photometric observations each day has revolutionized variable star research and enabled significant advances in asteroseismology. The high-cadence and long-baseline observations are opening new windows into the interior workings of stars and advancing our understanding of stellar structure and evolution \citep[see e.g.,][for comprehensive theoretical and observational overviews]{Aerts2010,Hekker2017,Bowman2020,Aerts2021,Kurtz2022,Bowman2023,AertsTkachenko2023}. With astronomical data sets reaching petabyte scales, automated analysis pipelines have become a key part of mission development that enable scientific discovery.

The NASA \textit{Kepler} mission \citep{Koch2010,Borucki2010} marked a milestone for stellar variability studies, delivering the uninterrupted light curves for approximately $160\,000$ stars at 30-min and 1-min cadence intervals for up to four years of time. Following the failure of its reaction wheel, the mission was repurposed into \textit{Kepler} Second Light \citep[K2, ][]{Howell2014} and observed stars across the ecliptic for 80 days per campaign. The NASA Transiting Exoplanet Survey Satellite \citep[TESS, ][]{Ricker2015} was launched in 2018 and is observing hundreds of millions of stars across the full sky, making it a treasure trove for data-driven science. TESS currently observes stars in sectors of 27.4 days, with total baselines spanning a few months to several years depending on sky position, and cadence ranging from 20-sec to 30-min, depending on mission cycle and data product. The ESA PLAnetary Transits and Oscillations of stars \citep[PLATO, ][]{Rauer2024} mission will be launched in 2026 and monitor the same patch of sky continuously for at least two years, following a similar strategy as \textit{Kepler}. The Gaia mission \citep{Gaia2016,GaiaDR3-2023} on the other hand, provides the sparse but very long-baseline photometric light curves for millions of stars, alongside its billions of high-precision astrometric measurements.

Ground-based surveys provide critical synergies with space missions because of their follow-up capacity and instrument capabilities difficult to obtain in space. Spectroscopic surveys such as the Sloan Digital Sky Survey V \citep[SDSS-V, ][]{Kollmeier2017} and GALactic Archaeology with HERMES \citep[GALAH, ][]{Buder2021} survey, provide detailed stellar parameters and chemical compositions, while photometric photometric surveys such as the All-Sky Automated Survey for Supernovae \citep[ASAS-SN, ][]{Shappee2014,Kochanek2017} and the Zwicky Transient Facility \citep[ZTF, ][]{Graham2019} provide long-baseline observations and follow-up capabilities. The upcoming Vera C. Rubin Observatory Legacy Survey of Space and Time \citep[LSST, ][]{Ivezic2019} will transform ground-based observing with its deep- and wide-field optical observations of over 10TB of data per night.

The combination of space- and ground-based data allows for the most accurate and detailed characterizations of stars and exoplanets \citep[e.g., ][]{Burssens2023,Hon2025}. However, discovering novel scientific insights from such datasets on a large scale is a challenge for traditional astronomical methods because they were developed for small samples of stars. Machine learning models on the other hand thrive with large datasets because their generalization ability scales with data size. From identifying similar groups of stars and inferring their parameters, to discovering rare astrophysical phenomena that require follow-up, machine learning is enabling the transition toward automated astronomical discovery.

This review focuses on the role of machine learning in variable star research and asteroseismology, with an emphasis on large space-based photometric surveys and their synergies with ground-based spectroscopic and photometric surveys. The review starts with an introduction to stellar variability and the instrumental effects impacting its characterizations, and is followed by an overview of variability classification and parameter inference techniques. This sets the stage for a discussion on the recent advancements in multimodal embeddings and foundation models, and the progress toward fully automated scientific discovery. As this review is part of the Springer Nature 2024 Astronomy Prize Awardees Collection, I pay particular attention to work I have personally been involved with and is closely aligned with my research.

\section{(Stellar) variability?}


Variable stars offer a unique window into the structure and evolution of stars. Their temporal brightness variations reveal information about their interior structure, surface layers, rotation rates, angular momentum transport and presence of planetary or stellar companions. Stellar variability can arise from changes in the physical conditions within the star itself (intrinsic variability), such as stellar pulsations, or from external influences (extrinsic variability), such as a companion star periodically obscuring part of the observed light in an eclipsing binary system. Stellar pulsations are of particular importance, as they allow us to probe the stellar interior \citep{Aerts2021}. Pulsations are small perturbations to a star's equilibrium state caused by different oscillation modes propagating through the stellar interior. Gravity- or g-modes have buoyancy as their main restoring restoring force and reach their largest amplitudes in the deep interior, making them ideal for studying the near-core region of stars. Pressure- or p-modes are acoustic waves that have pressure as their main restoring force and reach their largest amplitudes in the stellar envelope, making them ideal for studying the outer regions of stars. Binarity on the other hand can enable the determination of fundamental stellar parameters such as mass, radius, and orbital elements, while stellar rotation rates can enable age estimation through gyrochronology. The combination of these parameters across a diverse range and large sample of stars presents a unique opportunity for improving our understanding of stellar structure and evolution. This review primarily focuses on the TESS mission, as its full-sky coverage and high-cadence observations offer a particularly powerful tool for studying the stellar variability at an unprecedented scale.

Photometric stellar variability observations are often contaminated with non-astrophysical variability caused by various instrumental, systematic and crowding effects. While these effects impact small-scale studies, they pose even greater challenges for large-scale machine learning pipelines because of their sensitivity to training data and quality. Deviations from the assumption of independent and identically distributed (i.i.d.) data can significantly degrade the accuracy of machine learning algorithms. Systematic trends in astronomical data are not uniformly distributed, but rather correlated to observational properties such as the location of a star on the sky and CCD, or time of observation. Hence, having methods that can robustly correct for such effects is essential for ensuring reliable results.

\begin{figure*}[ht]
\centering
\includegraphics[width=1\textwidth]{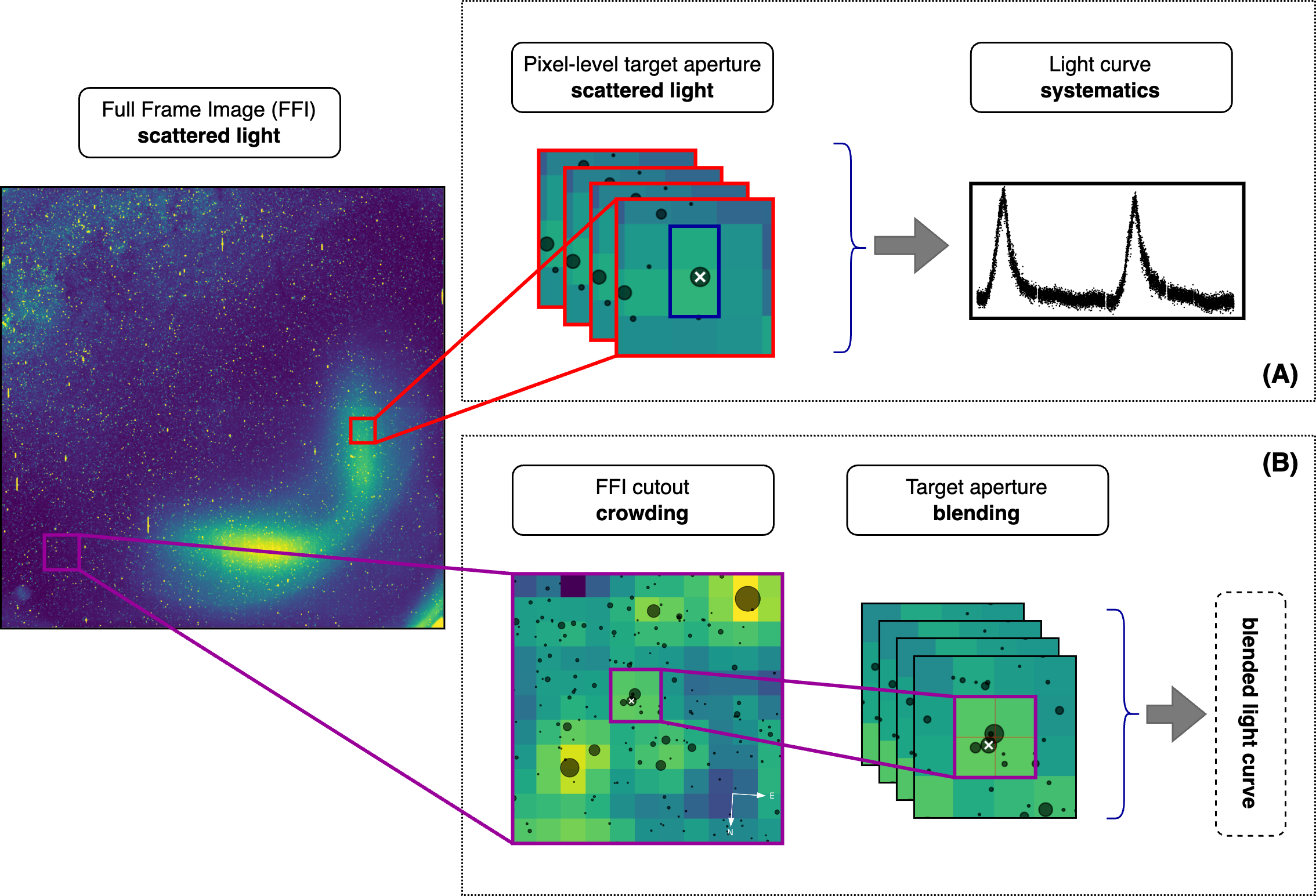}
\caption{Illustration of systematic and instrumental effects affecting TESS light curves. The top panel (A) shows the effect of scattered light on the light curve if not corrected. The bottom panel (B) shows the effect of blending in crowded fields due to the larger pixel size of TESS.}
\label{fig:systematics}
\end{figure*}

\subsection{Systematic corrections}

Systematic trends in photometric data can be addressed at different levels: the image level, light curve level, or a combination of both. For the TESS mission, different data reduction pipelines employ different approaches to mitigating instrumental and systematic trends. The most important pipelines are the MIT Quick-Look Pipeline \citep[QLP,][]{QLP1Huang2020a,QLP2Huang2020b,QLP3Kunimoto2021,QLP4Kunimoto2022}, TESS-Gaia Light Curves \citep[TGLC,][]{Han2023}, TESS Science Processing Operations Center \citep[SPOC,][]{Caldwell2020,Jenkins2016}, the methods from the TESS Asteroseismic Consortium \citep[TASC, ][]{Handberg2021,Lund2021} and \texttt{eleanor} \citep{Feinstein2019}. The main difference is that exoplanet-focused pipelines employ heavy filtering and binning strategies in order to more easily detect planets. This often removes astrophysical signals however, making the light curves ill-suited for stellar variability research. In this case, it is preferable to use the raw non-detrended light curves delivered by the pipeline in combination with custom filtering techniques that are optimized for retaining stellar variability signals. Leveraging the diversity of light curves produced by different pipelines, each with their distinct correction and aperture selection techniques, could serve as a strategy for mitigating instrumental effects.

One of the most visible systematic trends in TESS is scattered light from the Earth and the Moon entering the cameras\footnote{\url{https://tess.mit.edu/observations/scattered-light/}}, which manifests itself as significant increases in background levels that can obscure astrophysical signals. Panel (A) of Fig.~\ref{fig:systematics} illustrates how scattered light in the Full Frame Image (FFIs) introduces strong trends into the extracted light curve, masking the stellar variability signal. Since scattered light is a function of the angle and distance of the satellite to the Earth and the Moon, and to camera and CCD location, corrections are ideally performed at the image level.

While mitigation strategies are part of the data reduction pipelines that were mentioned earlier, they are often imperfect. \citet{MuthukrishnaLupo2024,Lupo2024} are therefore approaching the corrections in a data-driven way through the development of a deep learning methodology that learns to correct the TESS Full Frame Images for scattered light. Their methodology leverages conditional diffusion models \citep{ho2020} to probabilistically model the dynamic scattered light patterns, enabling the prediction of background levels for future sectors that can be subtracted from the raw images. This produces corrected Full-Frame Images free of scattered light together with associated uncertainties, providing clean photometric images for light curve extraction.

On the light curve level, techniques based on Principal Component Analysis (PCA) have been widely used to correct for systematic effects \citep[e.g.,][]{Smithkepdatc8,Jenkins2016}. Recently, more sophisticated methods have emerged. For example, \citet{Hattori2022} used half-sibling regression \citep{Scholkopf2016}, a technique that leverages the causal structure of data, to detrend individual pixel light curves from scattered light and common systematics, achieving improvements without requiring full image and light curve reprocessing of the millions of TESS observations.

\subsection{Blending and contamination}

Blending occurs when the light of multiple stars falls onto the same pixels or target aperture. This leads to a light curve in which the signal of one star is contaminated by another, which can lead to false exoplanet detections or inaccurate variability classifications. The issue is particularly prevalent in TESS because its relatively large pixel size of 21 arcec increases the likelihood of observing blended sources, especially in crowded fields. Panel (B) of Fig.~\ref{fig:systematics} shows an example where three stars of comparable magnitude fall in the same light curve aperture, complicating the analysis of the light curve.

Tools such as \verb+TESS_localize+ \citep{HigginsBell2023} have been developed to analyze light curve blending. By leveraging the differences in variability patterns between different stars, the origin of the variability on the sky can be localized. \citet{PedersenBell2023} demonstrated the utility of this method by showing that a newly claimed type of variability was, in reality, the result of blending. Properly accounting for blending through contamination metrics is essential for automating variability analysis, as inaccurate source attribution of variability can lead to incorrect estimations of stellar properties, incorrect variability classes and anomalies that are not astrophysical in origin.

\section{Variable star classification}

The exponential growth in astronomical observations has created massive data sets of stars with different characteristics spanning the full breadth of physical properties. This diversity in variability types and data size requires efficient classification algorithms that structure the incoming data according to its physical characteristics. 

Machine learning algorithms are ideal for this task because of their ability to learn complex and similar patterns in large datasets. Whether to use supervised learning, which relies on labeled data to train a classification model, or unsupervised learning, which relies on the underlying structure and distribution of data to find clusters, depends on the research objectives. Supervised methods excel at rapidly classifying observations according to current astrophysical knowledge, but come at the expense of a labeling bias. Unsupervised methods on the other hand, do not impose a human bias. This makes it possible to discover new clusters of stars and potentially reveal previously unknown physical correlations. However, it comes at the expense of efficiently classifying stars according to known predefined variability types.

\begin{figure*}[ht]
\centering
\includegraphics[width=1\textwidth]{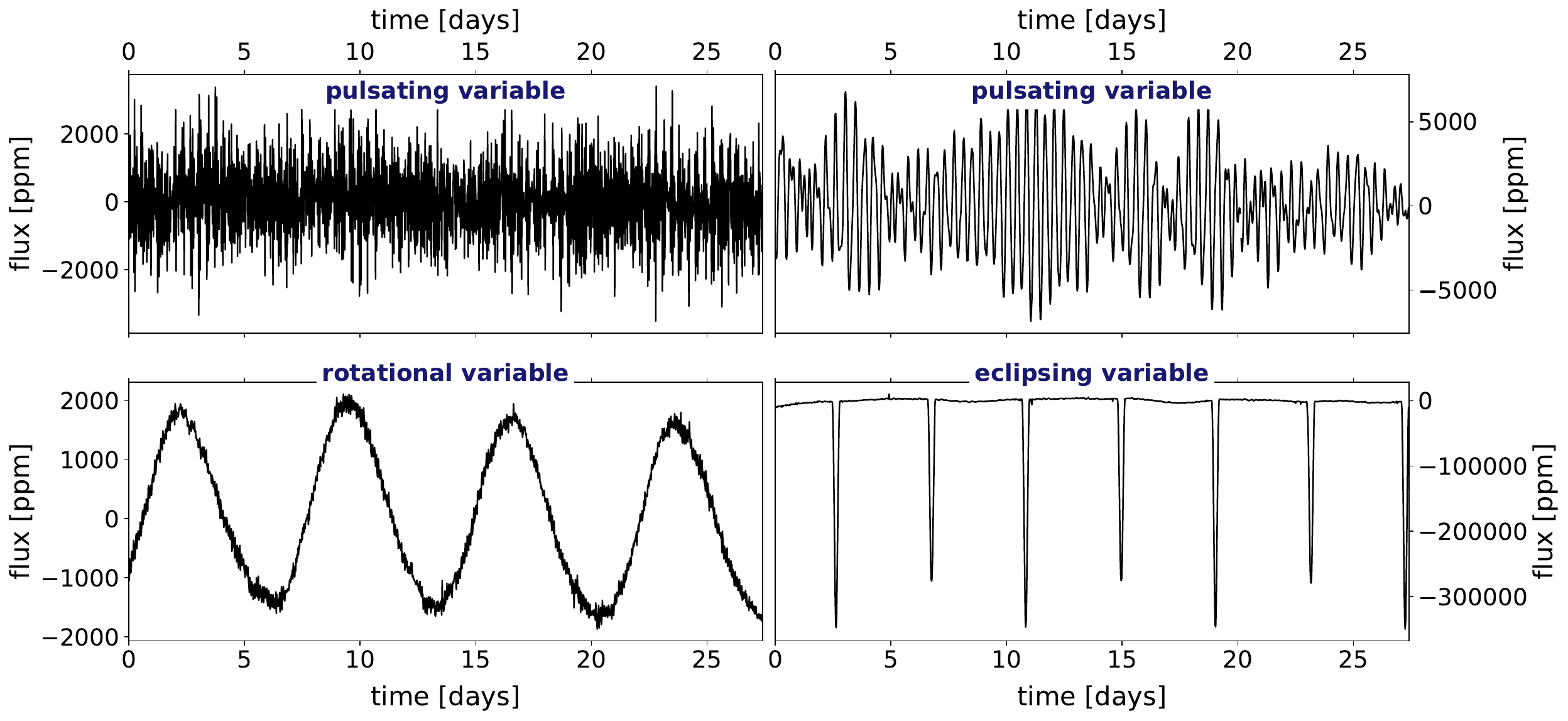}
\caption{Example light curves of four variable stars: two pulsating variables, a rotational variable and an eclipsing variable.}
\label{fig:lc}
\end{figure*}

In supervised classification, variability classes can, for example, be structured according to the variability tree presented by \citet{EyerMowlavi2008,Eyer2019}, depending on the available input data (photometric, spectroscopic,...). Fig.~\ref{fig:lc} shows four example light curves of stars with different variability types. The classifications can serve as the basis for follow-up studies \citep[e.g.,][]{Hon2021} or deciding follow-up observations \citep[e.g.,][]{Muthukrishna2022}.  Unsupervised learning can offer new insights into how stars transition across the Hertzsprung-Russel Diagram. For example, \citet{Audenaert2022} demonstrated the potential of clustering hybrid pulsators with both pressure (p-) and gravity (g-) modes \citep[see e.g.,][]{Uytterhoeven2011,Bowman2018} using biomedical data processing techniques, and for probing rotational properties. A hybrid approach in which light curves are first classified with supervised learning according to established variability classes, and then clustered in detail using unsupervised learning can provide the best of both worlds: efficient categorization while still allowing for a more detailed unbiased exploration.

\subsection{Machine learning}

The high dimensionality, instrumental characteristics and noise properties of light curves pose challenges for directly feeding them into automated classification models. Machine learning models therefore either rely on feature engineering or feature learning to transform light curves into lower dimensional representations that capture their characteristic variability properties. In the case of feature engineering, features are constructed manually based on domain expertise, whereas in the case of feature learning or representation learning, a model automatically learns an informative representation based on a certain objective. The constructed features or learned representations form the basis for a model to learn the decision boundaries of the different variability classes.

\paragraph{Feature engineering: leveraging domain expertise}
Light curve variability classifiers in astronomy have long relied on feature engineering-based models. Random forest classifiers \citep{Breiman:fb}, which rely on ensembles of decision trees trained on sets of input features, have proven to be one of the most robust and easy to train classification models. Their ensemble architecture makes them less prone to overfitting compared to neural networks, particularly when combined with careful feature engineering. The construction of input features for such types of models has mostly been guided by traditional astrophysical and statistical methods, with features based on the frequency spectrum \citep[e.g., Lomb-Scargle periodogram,][]{lomb1976,scargle1982}, moments (e.g., skewness) or entropy \citep{Shannon1948}. These transformations reduce the dimensionality of the light curve and are an effective way to reduce noise and improve signal detection, while at the same time improving robustness for heterogeneous data.

The basis of feature-based discriminative stellar variability classifiers based on light curve information was laid by \citet{Debosscher2007,Debosscher2009,Debosscher2011,Sarro2009b,Blomme2010,Blomme2011,Richards2011}, with more recent examples being \citet{Kim2016,Kuszlewicz2020,Barbara2022}. In a step to automatically learn the representative features from a light curve, \citet{Armstrong2016} used Self-Organizing Maps \citep[SOM,][]{Kohonen1990}, following earlier work by \citet{Brett2004}.

\paragraph{Representation learning: learning data-driven representations}
Representation learning, the core idea behind artificial neural networks, aims to automatically discover the characteristic representations of data, alleviating the need for time-consuming manual feature engineering. By not having a manual bias, the algorithm learns to pick up on the characteristics important for the maximization of its objective function (e.g., minimizing misclassifications by minimizing the difference between predicted and target labels). Feed-forward neural networks \citep[see e.g,][]{Goodfellow2016} achieve this by learning internal representations encoded in a set of weights through which the input data is transformed to produce output labels.

Training feed-forward neural networks directly on full light curves is particularly challenging due to the broad range of variability timescales (e.g., minutes to days), high dimensionality and heterogeneous noise properties. Additionally, feed-forward networks require fixed-length input and output vectors, whereas light curves mostly have varying lengths, necessitating pre-processing steps.

Many deep learning architectures in astronomy therefore rely on phase-folded light curves or frequency spectra as input rather than on the full light curve, which requires prior period estimation and Fourier transforms. \citet{Hon2018a,Hon2018b,Hon2019} used the power density spectra of light curves as input to a Convolutional Neural Network \citep[CNN, see e.g.,][]{LeCun1989,Lecun1995,LeCun1998,LeCun2015}, while \citet{Cui2022,Cui2024} implemented automated zooming algorithms to allow the CNN to more effectively analyze light curves and their power spectra with different scales, mimicking human-like classification strategies. Although CNNs learn higher-level representations of the input data, the discussed models rely on pre-transformed versions of a light curve (e.g., Fourier transform or phase-folded light curve) instead of on the raw light curve, making them hybrid approaches that combine feature engineering with some form of feature learning.

In order to create a robust machine learning model optimized for classifying the millions of \textit{Kepler} and TESS light curves, the TESS Data for Asteroseismology (T'DA) working group within the TESS Asteroseismic Science Consortium (TASC) developed an ensemble model that combines feature engineering with feature learning based approaches \citep{Audenaert2021}. The architecture uses stacked generalization \citep{Wolpert1992} to combine the predictions of random forest, gradient boosting \citep{friedman2001} and CNN \citep{LeCun2015} classifiers that are each trained on different feature sets, such as the period, entropy and power density spectrum of a light curve. In this way, the ensemble model can account for the relative strengths of each of the individual classifiers and achieve a higher overall accuracy, while at the same time improving the robustness of its predictions compared to any single classifier. One of the key legacies of this collaborative effort by the asteroseismology community is the creation of a high-quality training set of light curves, which now serves as a valuable benchmark for developing and validating future stellar variability classifiers.

Other variability classifiers based on light curve information include the work by \citet{Mahabal2017,Sanchez-Saez2021,Olmschenk2024}, with examples in exoplanet detection by \citet{Olmschenk2021,Tey2023}. Additionally, the usage of anomaly detection techniques can also be valuable for classification, as the variability characteristics of rare classes could drive anomaly scores \citep[e.g.,][]{MartinezGalarza2021,Crake2023}.

\paragraph{Time-series representation learning: deep models for sequential data}

Feed-forward neural networks were not specifically designed for processing sequential data, such as light curves. Recurrent Neural Networks \citep[RNNs,][]{Rumelhart1986} and Long Short-Term Memory Networks \citep[LSTMs,][]{Hochreiter1997} by contrast, were developed for this purpose and can handle variable length sequences \citep{Cho2014,Sutskever2014}, making them better suited for automatically learning representations of light curves in an end-to-end fashion. \citet{naul2018,Becker2020,jamal2020} laid the basis for using these types of deep learning architectures for stellar variability classification.

However, despite the conceptual advantages of these models, achieving a performance that rivals the robustness of random forest, feature-based and ensemble classifiers in a production context has long been been challenging. Traditional RNNs suffer from the vanishing gradient problem, making them difficult to train. More recently, the introduction of the attention mechanisms in transformer architectures \citep{vaswani2017} has led to significant performance improvements, because of their ability to better take into account long-range dependencies and their greater parallelizability. \citet{Cabrera-Vives2024} demonstrated that transformers outperform traditional feature-based random forest classifiers in the classification of simulated LSST data, while Gregory et al. (in prep.) are finding similar results for \textit{Kepler} and TESS light curves.

\subsubsection{Domain adaptation}

One challenge in deploying variability classifiers across the field is that models are often developed and trained for specific surveys and instruments. The observations from different telescopes can be very different, however. The uninterrupted high-cadence observations from space (e.g., Kepler or TESS) significantly differ from their ground-based counterparts (e.g., LSST), which contain daily gaps, are irregularly sampled and sparse but can have very long baselines. This can cause the model weights to be misaligned for the target domain. Transfer learning and domain adaptation techniques have proven to be valuable in adapting models to another survey \citep[e.g.,][]{Kim2021,Ciprijanovic2023,GuptaMuthukrishna2025}, allowing easier portability and generalizability of a model across data sets. However, if the data are too different, the architectures can be incompatible because they, for example, require a minimum number of samples per input and cannot deal with sparse light curves, or the opposite, they cannot deal with light curves consisting of a very large number of samples. 

\subsubsection{Latency considerations}

The latency requirements for classification are an important factor to consider when developing machine learning architectures and depend on the science case \citep[see e.g.,][Fig. 1]{Agarwal2023}. For example, supernovae and transients require rapid classification in order to enable timely follow-up observations. \citet{Muthukrishna2019a,Muthukrishna2019b,Muthukrishna2022} therefore developed real-time transient detection pipelines using deep learning for large time-domain surveys. In contrast, pulsating stars benefit from long-baseline observations for their characterization and hence do not require a low-latency classifier.

\subsection{Synergies with citizen science and statistics}

The analysis of large data volumes can be also approached with techniques other than machine learning. Citizen science initiatives in particular have proven to be very successful, as demonstrated by \citet{Eisner2021} for the detection of TESS planet candidates \citep[e.g.,][]{Eisner2024}. The citizen science results could be coupled with machine learning to achieve the best of both worlds, as shown by \citet{Tardugno2024}, who used a CNN to filter false positives from the Planet Hunters TESS data set. 

Statistical and mathematical methodologies are important for creating high-quality variability catalogs \citep[e.g., ][]{Prsa2022}, as supervised machine learning algorithms rely on previously labeled samples for training. For example, \citet{IJspeert2021,IJspeert2024a,IJspeert2024b} leveraged the derivatives of light curves in combination with statistical modeling techniques to create an automated pipeline for the detection of eclipsing binaries.

Despite their strengths, citizen science and purely statistically-driven methodologies can face limitations when dealing with large numbers of variability types. The complexity of training humans to distinguish between a large number of variability types can make it challenging for citizen science, while the specificity of statistical models to a particular variability class can make it challenging to generalize the model for other classes with different characteristics. Machine learning algorithms, in contrast, offer greater flexibility and scalability, as they can more easily be trained to classify different variability classes and rapidly classify millions of observations.

The combination of different methodologies helps mitigate biases and leads to more comprehensive and accurate catalogs. By integrating data from different surveys and by combining machine learning with statistical and visual classifications, such as information from TESS \cite[e.g.,][]{Fetherolf2023,Skarka2022,Skarka2024,Nielsen2022,Hatt2023}, Gaia \citep[e.g.,][]{Rimoldini2023,Eyer2023,DeRidder2023}, TESS and Gaia \citep[e.g.,][]{Hey2024} or ASAS-SN \citep[e.g.,][]{Jayasinghe2018, Jayasinghe2019}, robust catalogs can be constructed. Ultimately, the most promising approach lies in the creation of multimodal embeddings that directly integrate the data from different instruments and surveys into one model, as will be discussed in Sect.~\ref{Sect:FM}.

\section{Stellar parameter inference}

The homogeneous samples of variable stars obtained through classification provide a base for more detailed astrophysical modeling efforts. Stellar parameter inference is closely intertwined with classification, as stars with similar characteristics should ideally be grouped together. However, supervised classification methods can introduce a human bias in this sense as they rely on predefined variability classes that may overlook certain hidden patterns and relationships within the data. Fully data-driven clustering methods could reveal new insights into stellar parameters, as stars can be structured according to their underlying properties rather than according to predefined classes.

\subsection{From clustering to stellar parameters}

The lower dimensional representations of light curves can reveal a wealth of information on the physical characteristics of stars. \citet{Audenaert2022} developed an interpretable clustering methodology to tackle the case of variable stars with overlapping properties. P-mode and g-mode pulsators of spectral types F and A, such as $\delta$\,Scuti and $\gamma$\,Doradus stars \citep[see e.g.,][]{Aerts2010}, have partially overlapping instability regions in the Hertzsprung–Russell Diagram \citep{Dupret2004}, giving rise to a class of hybrid pulsators with both p- and g-modes \citep{Dupret2005,Uytterhoeven2011,Bowman2016}. From an astrophysical perspective, hybrid pulsators are of particular interest because the presence of both p- and g-modes, respectively, allows to probe their envelope and near-core properties simultaneously.

The clustering methodology from \citet{Audenaert2022} is rooted in biomedical science and leverages Electrocardiogram (ECG) heart-rate variability signal processing techniques as the basis for light curve characterization or feature engineering. The multiscale entropy \citep{Costa2002,Costa2005} captures complexity of a light curve and is used as input for a clustering analysis with UMAP \citep[Uniform Manifold Approximation and Projection,][]{McInnes2018,mcinnes2018umap-software} and HDBSCAN \citep[Hierarchical Density-Based Spatial Clustering of Applications with Noise,][]{McInnes2017,Campello2013,Narayan2021}. The methodology can distinguish pure p- and g-mode pulsators from their hybrid counterparts and allows for new insights into the transition regions. One of the most exciting findings is the correlation between latent space structure and stellar parameters and in particular with rotation, which is essential for understanding angular momentum transport \citep[e.g.,][]{Aerts2019,AertsTkachenko2023,Mombarg2023}.

\begin{figure*}[ht]
\centering
\includegraphics[width=1\textwidth]{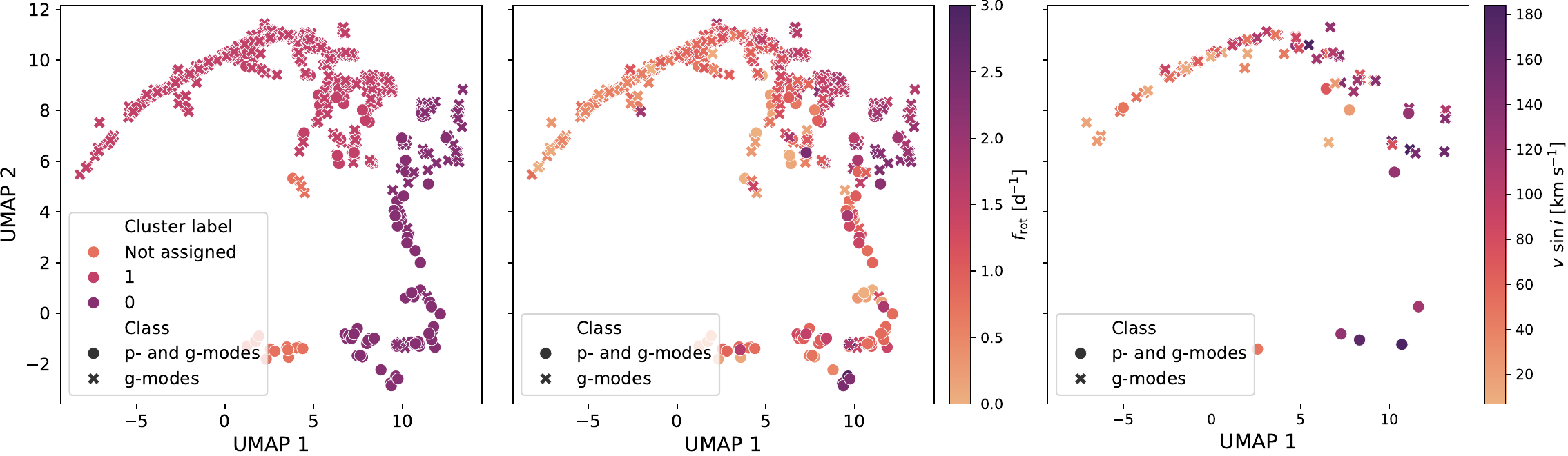}
\caption{Clustering structure of the pure g-mode and hybrid p- and g-mode pulsators in the $\gamma$\,Doradus catalog from \citet{Li2020}. From left to right panel, the data is color-coded according to the clusters found with HDBSCAN \citep[from][]{Audenaert2022}, near-core rotation rate $f_{\rm rot}$ \citep[from][]{Li2020} and spectroscopic $v \sin i$ \citep[from][]{Gebruers2021}. The pulsation class labels are indicated by the different marker shapes and were assigned by \citet{Li2020} based on a visual inspection of the light curves and power spectra. The methodology from \citet{Audenaert2022} is unsupervised so the plotted labels were only used for validation and not for training. Figure partially reproduced from \citet{Audenaert2022}.
}
\label{fig:clustering}
\end{figure*}

This is illustrated in Fig.~\ref{fig:clustering}, which shows the two-dimensional UMAP representation of the multiscale entropy for the sample of pure g-mode and hybrid p- and g-mode pulsators from \citet{Li2020}. The plots are, from left to right, color-coded by the cluster identified with HDBSCAN, asteroseismic near-core rotation rate \citep[from][]{Li2020} and spectroscopic surface rotation rate \citep[from][]{Gebruers2021}, while the marker shape shows the class assigned by \citet{Li2020} based on a visual inspection. The structure of the data can clearly be explained by the pulsation modes and rotation rates, demonstrating the potential of estimating stellar properties without depending on labeled training data. It also shows the potential for leveraging multimodal data to better understand differential rotation by comparing surface and near-core rotation rates.

\subsection{Stellar modeling with machine learning}

The previous section showed the potential of capturing the properties with data-driven methods. Stellar parameters can be inferred efficiently on a large scale by training supervised machine learning models on the results from existing statistical and stellar models. The machine learning models effectively act as function approximators, learning the relationships between observational data and theoretical physical parameters. Rotation periods can be inferred from TESS and \textit{Kepler} light curves with rotational modulation \citep[e.g.,][]{Lu2020,Breton2021,Claytor2022,Colman2024}. In this case, the models learn to approximate the output of traditional mathematical period estimation algorithms such as periodograms, autocorrelation functions \citep[e.g., as in][]{McQuillan2014} and wavelet analyses.

The benefit of machine learning lies in its ability to filter out irrelevant variability components, leading to more accurate rotation estimates, while at the same time reducing the computational complexity compared to traditional methods. The latter is an essential factor in moving toward large-scale studies based on surveys such as TESS.

Similarly, the automated analysis of spectroscopic data, such as from the Sloan Digital Sky Survey (SDSS),
enables a fast determination of stellar labels such as effective temperature, surface gravity and metallicity \citep[e.g.,][]{Ness2015,OBriain2021,Ting2019,Straumit2022}. Additionally, machine learning models achieve higher consistency and precision compared to manual label fitting methods by reducing individual biases and enabling more reproducible parameter determinations, with ongoing efforts to better incorporate uncertainties \citep[e.g.,][]{Horta2025}. Some of the stellar labels, such as the surface gravity, can also be determined from photometric data \citep[see e.g.,][]{Bugnet2018,Ness2018,Sayeed2021,Pan2024}, illustrating the potential for leveraging correlations between multiple modalities.

The emulation of stellar structure and evolution models, and physical models in general, with machine learning techniques \citep[e.g.,][]{Hendriks2019,Mombarg2021,Scutt2023,Hon2024,Mombarg2024,Wrona2024} allows to more quickly estimate the theoretical photometric or spectroscopic observables that can be used to constrain actual observations, by reducing the need for extensive physical simulations each time. At the same time, emulators can make it computationally tractable to create higher dimensional models. By systematically varying the set of physical variables in grid-based simulations and training a machine learning model on the resulting output, the model can learn to interpolate between different variables. The effectiveness does depend on the ability of the model to reliably interpolate between the sparse data grids and generalize it to the entire parameter space. Incorporating a broad set of physical input variables helps reduce uncertainties and facilitates a deeper investigation of complex stellar physics.

\section{Foundation models}
\label{Sect:FM}

The advancements in classification and parameter inference methods have significantly contributed to stellar variability and asteroseismology research. By enabling the identification of large homogeneous samples of stars and the determination of their fundamental properties, observational findings can be confronted with theoretical models, revealing their discrepancies and driving new astrophysical insights. However, while current machine learning tools provide powerful analysis tools, they are not directly probing the underlying physics of the observed phenomena.  

In order to create models capable of learning the underlying properties of stars, we need self-supervised techniques that can learn unified representations based on multiple data modalities in a label-independent manner. The embeddings learned by self-supervised models serve as lower dimensional representations of the observations capturing their physical properties. The learned latent space forms the foundation for a wide range of downstream tasks, ranging from clustering to parameter inference. The foundation models \citep{bommasani2021} built on these embeddings then alleviate the need of training supervised models from scratch for every dataset or scientific question.

Unlike traditional approaches that rely on precomputed statistical and mathematical summary statistics, self-supervised learning techniques rely on surrogate tasks to learn representations of the data. One example is masking parts of the input data and training the model to predict missing parts. Alternatively, models can be trained to differentiate between similar and dissimilar data points with contrastive learning in order to learn to extract representations. In this case, augmentation strategies can be used to create the similar data points \citep[see e.g., simCLR,][]{Chen2020}. Models can also be trained to learn to align the different data modalities for the same underlying target \citep[see e.g., CLIP, ][]{radford2021}. In this case, the image of a star and the word ``star'' could be forced to be close to each other in the latent space. Overall, these techniques allow models to learn in a self-guided manner, extracting meaningful information without explicit labels, making them suitable for different downstream tasks.

The first steps toward the development of foundation models for astronomy are currently being made. Recent works, such as \citet{Parker2024,Rizhko2024,Zhang2024,Koblischke2024,Huijse2025}, have explored various self-supervised learning techniques for creating embeddings of multimodal astronomical datasets that can be used as the basis for downstream tasks, with some broader galaxy and physics examples being \citet{Walmsley2022,Smith2024,EuclidSiudek2025,Birk2024,Hallin2024}. The work by \citet{Pimentel2023,Donoso-Oliva2023,Donoso-Oliva2025} on the other hand, focuses using on self-supervised transformer-based architectures to create light curve representations and can serve as a basis for the time series representation component in multimodal models.

One particularly exciting development is causal representation learning. \citet{audenaertmuth2025} are leveraging the causal structure of astronomical observations to learn more accurate physical embeddings as the basis for a foundation model for variable stars. By utilizing overlapping observations across the NASA Kepler and TESS missions, with the potential to integrate other data types such as spectroscopy, in a contrastive learning framework, the instrumental properties of the data can be separated from the physical properties, laying the basis for an interpretable physical foundation model.

The successful development of foundation models in astronomy requires access to large, diverse and multimodal datasets that span the full breadth of astronomical observations. A significant initiative in this regard is The Multimodal Universe \citep{mmu2024}, a large 100TB multimodal data set with astronomical observations ranging from multivariate time series to hyperspectral images. This is a key step in bringing machine learning and astronomy closer together and in enabling the creation of foundation models.

\section{Concluding remarks}

In this invited review, I have discussed the exciting progress that is being made in leveraging machine learning to obtain new astrophysical insights. The ever-growing amount of astronomical data has made automated data processing and analysis techniques a key factor in enabling astronomical discovery. After highlighting the importance of input data quality, I discussed how machine learning classifiers have transformed variable star research with their ability to rapidly mine millions of light curves in search for interesting targets. Traditionally, machine learning architectures relied on feature engineering to extract the characteristics properties of light curves. While it has long been challenging to rival the robustness of feature-based classifiers, we are now moving to an era where representation learning-based methodologies that automatically learn the data properties without manual intervention are surpassing the performance of feature-based architectures. By learning the representations with self-supervised techniques, they can form the basis for a variety of different tasks, from identifying different types of pulsating stars to inferring their parameters. The novel development of multimodal representations that integrate multiple data sources (e.g., photometric light curves, spectroscopy,...) is proving particularly promising in our quest to capture the underlying physical physical characteristics of stars. The multimodal embeddings can serve as the basis for foundation models that can be used for downstream tasks across surveys and instruments.

The results provided by machine learning classifiers not only facilitate the selection of which stars to study in more detail, but can also help to inform the observing strategies of other telescopes. The detection of interesting targets can trigger follow-up observations with other telescopes, while large-scale variability catalogs can help in planning future space missions. For example, the PLATO Field-of-View can be explored through TESS observations \citep[e.g.,][]{Eschen2024}, where TESS variability classifications could be used to select the optimal targets to observe with PLATO.

While machine learning offers great promises, it is important to note that it should not be used for every possible task as explained by \citet{Hogg2024}. In particular, the combination of multiple separate catalogs from different machine learning algorithms does necessarily create an unbiased sample that is a reliable estimate for population studies. While every survey inherently has selection biases due to instrumental limitations and observing strategies, machine learning models have inductive biases that are not always well understood, potentially skewing their output even when appearing unbiased. Moreover, machine learning models often lack reliable uncertainty quantification, which is essential for understanding the reliability of their findings. Incorporating robust uncertainty estimation techniques will significantly enhance the utility of machine learning results for cross-survey comparisons and astrophysical research.

Looking ahead, the creation of multimodal embeddings that integrate data from different surveys while specifically accounting for the underlying data quality of each observing instrument is especially promising, as aligning data with varying observational properties remains a significant challenge. The development of these models offers the potential to seamlessly combine data from a wide range of sources, including from small telescopes, such as the Mercator telescope HERMES spectrograph \citep{Raskin2011,Royer2024}, the currently being commissioned MARVEL radial-velocity spectrograph \citep{Raskin2020}, and even cubesats \citep[e.g., CubeSpec,][]{Bowman2022Cube}. By unifying all available modalities within a foundation model that accounts for their systematic differences, we move closer to creating a comprehensive framework for astrophysical discovery that will be capable of unraveling the complexities of the universe through data-driven science.

\backmatter

\bmhead{Acknowledgements}
Funding for the TESS, Kepler and K2 mission is provided by NASA's Science Mission Directorate. The author would like to thank the reviewer, DM, DB and AT for their valuable feedback regarding the manuscript.

\bmhead{Ethics declarations} Not applicable.









\bibliography{sn-bibliography}

\end{document}